\documentclass[12pt,a4]{article}

\usepackage{amsmath}

\usepackage{amsfonts}
\usepackage{amssymb}
\usepackage{graphicx}

\def\be{\begin{eqnarray}}
\def\ee{\end{eqnarray}}

\def\nn{\nonumber}

\newcommand{\mpl}{M_{\rm Pl}}

\def\pa {\partial}
 
 \def\ra {\rightarrow}
 
 \def\neq {\not\equiv}

\def\cs2{c_{s}^{2}}

 \def\be   {\begin{equation}}   \def\ee   {\end{equation}}
 \def\ba   {\begin{array}}      \def\ea   {\end{array}}
 \def\bea  {\begin{eqnarray}}   \def\eea  {\end{eqnarray}}
 \def\bean {\begin{eqnarray*}}  \def\eean {\end{eqnarray*}}

\begin{document}

\begin{flushright} {\footnotesize CERN-PH-TH/2011-56}  \end{flushright}
\vspace{5mm}
\vspace{0.5cm}
\begin{center}
\def\thefootnote{\fnsymbol{footnote}}

{\Large \bf The Kramers-Moyal Equation 
\\ [0.3cm]
of the Cosmological Comoving Curvature Perturbation
}
\\[0.5cm]
{\large Antonio Riotto$^{\rm a,b}$ and Martin S. Sloth$^{\rm a}$}
\\[0.5cm]

\vspace{.2cm}

{\small \textit{$^{\rm a}$  CERN, PH-TH Division, CH-1211, 
Gen\`eve 23,  Switzerland}}

\vspace{.2cm}

{\small \textit{$^{\rm b}$  INFN Sezione di Padova, via Marzolo 8,
I-35131 Padova, Italy}}

\end{center}

\vspace{2.5cm}

\hrule \vspace{0.3cm}
{\small  \noindent \textbf{Abstract} \\[0.3cm]
Fluctuations of the comoving curvature perturbation with wavelengths larger than the 
horizon length are governed by a  Langevin equation whose 
stochastic noise  arise from the quantum fluctuations that are assumed to become classical at horizon crossing.
The infrared part of the curvature perturbation  performs a random walk under the action of the stochastic noise
and, at the same time, it suffers a classical force caused by its self-interaction. By a path-interal approach and, alternatively, 
by the standard procedure in random walk analysis of adiabatic elimination of fast variables, 
we derive the corresponding 
Kramers-Moyal  equation which describes how the probability distribution of the comoving curvature
perturbation   at a given spatial point evolves in time and is a generalization of the Fokker-Planck equation. 
This  approach 
offers  an alternative way to study the late time behaviour of the correlators
of the curvature perturbation from infrared effects.

\vspace{0.5cm}  \hrule
\newpage
\def\thefootnote{\arabic{footnote}}
\setcounter{footnote}{0}

\section{Introduction}
\noindent
Cosmological inflation \cite{lr} has become the dominant paradigm
within which one can attempt to
understand the initial conditions for Cosmic Microwave Background (CMB)
anisotropies and structure formation. In the inflationary picture, 
the primordial cosmological perturbations are created from quantum 
fluctuations which are
``redshifted'' out of the horizon during an early period of 
accelerated expansion.
Once outside the horizon, they remain ``frozen'' until the horizon
grows during a later matter- or radiation-dominated era.
After falling back inside the horizon they are communicated to the primordial
plasma and hence are directly
observable as temperature anisotropies in the CMB.
These anisotropies have been  recently  mapped 
with spectacular accuracy by the  Wilkinson Microwave 
Anisotropy Probe (WMAP)~\cite{wmap7}.

These CMB observations show that the cosmological perturbations are very small,
of order $10^{-5}$ compared to the homogeneous background.
Therefore, one might think that first-order perturbation theory
will be adequate for all comparison with observations.
However, that may not be
the case; the Planck satellite \cite{planck} and its
successors may be sensitive to non-Gaussianity  in the cosmological
perturbations.
Such non-Gaussianities are sourced by self-interactions in the early universe,
and become visible
at the level of second- or higher-order perturbation theory \cite{review}.
This possibility is of considerable interest in its own right
and is presently being vigorously explored.
However, there are other equally compelling reasons to go beyond linear
theory.
Self-interactions of any scalar field during the inflationary stage, and
more interestingly of the comoving curvature perturbation
$\zeta$,  give rise also to  corrections in all correlators,
including the observationally interesting cases of the
two- and the three-point correlation functions.
Such  corrections are associated with so-called ``loops,''
in which virtual particles with arbitrary momentum are emitted and
re-absorbed by the fields which participate in the correlation function. 
Loop corrections to the correlators of cosmological perturbations during inflation
have been extensively studied in the last few years
\cite{loops1,loops2,loops3,loops4,loops5,loops6, loops7}.
In particular,  it has been known for a long time that there exists a breakdown
in the perturbative
expansion due to infra-red (IR) divergences  for the case of an interacting 
 test scalar field
in a pure de Sitter stage \cite{sasaki1,sasaki2}.  Recently the IR issue was anlyzed in the more interesting (but also more challenging) case of real metric perturbations, taking also into account the fluctuations of the background, and it was shown that in this case there appears to be a perturbative breakdown globally in the "large box" on a time scale, $t\sim RS$, given by the de Sitter radius and entropy \cite{loops7}, which indicates some interesting connections to black hole physics and the information paradox \cite{bhp1,bhp2,bhp3,bhp4}. 

The  semi-classical 
one-loop IR corrections depend on the (logarthmic value of)  infra-red comoving momentum cut-off $L^{-1}$.
and there 
has been quite a debate in the literature on the interpretation of these
IR corrections \cite{ir1,ir2,ir21,ir3,ir4,ir401,ir402,ir41,ir5,ir61,ir7,ir8,ir9,ir10}.

One option will be to restrict ourselves
to the computation of cosmological observables  within our local patch. This means that 
all computations should be done within a comoving box of size $L$
whose present size is not much larger than the present horizon
$H_0^{-1}$, in order that the answer depend only on local properties
of the universe at our position. In such a case, IR corrections are of the order of $\ln(kL)={\cal O}(1)$ if we
observe the properties of the cosmological correlators on a given observationally  interesting 
comoving wavenumber $k$. 

The other option
is to set  $L$ to be some comoving length scale which left the
horizon many e-folds before the observable universe. In particular,
the smallest possible value of $L^{-1}$ is  
$a_{\rm i} H$, where $a_{\rm i}$ is the value of the scale factor at the beginning
of inflation. Since the wavelength
$k^{-1}$ goes out of the horizon when the scale factor equals
$a_k=k/H$, in such a case  
$\ln(k L)=\ln(a_k/a_i)$ is proportional to the total
number of e-folds from the beginning of inflation to the time when the
mode $k$ exits the horizon. Nevertheless, these large IR corrections appear only in
quantities which are not directly observable \cite{ir2,ir3} today. Indeed, 
one should be suspicious that large effects from
IR modes are somehow unphysical for a present observer, because on general grounds
we expect the structure within our observable patch of the
universe to depend only on its local properties.
Although their appearance cannot be avoided, they do not themselves have
any particular interest unless one wishes to ask more general 
questions, for instance, what is the probability to find an 
inflaton field homogeneous enough to 
lead to the correct temperature anisotropy in the CMB within patches like ours? Or,
what is the probability that the power spectrum of $\zeta$ has the value we measure? If one insists in adopting a superlarge box, it should be kept in mind
therefore that the location of our box
may be untypical and that the large IR effects appear as a measure of how likely it is that the correlators
averaged over the superlarge box coincide with the correlators in
our observable universe. 
One may also be interested in understanding the perturbative breakdown in the large box,
because it teaches us something fundamental about de Sitter space-times, which can be valuable for understanding also other problems. In any case, there appears to be a growth of perturbations in de Sitter, and we are interested in properly understanding its physics. 

To deal with this problem, but also for other less ambitious reasons,
 one can try to use the approach  of stochatic
inflation to study the behaviour of the IR part of the comoving curvature perturbation $\zeta$. 
The stochastic approach  embodies the idea that the IR part of the  field may 
be considered as a classical space-dependent
stochastic field satisfying a local Langevin-like equation \cite{starob}. 
The stochastic noise term arises from the quantum fluctuations which become
classical at horizon crossing and then contribute to the background. 
The IR part of the field  $\zeta$ performs a random walk under the action of this stochastic noise
and, at the same time, it suffers a classical force caused by its self-interaction. Physically, the stochastic  force is caused by the ultraviolet modes that are instantaneously redshifting across the horizon at time $t$. It is Markovian  because each increment is suitably  chosen to be uncorrelated with the previous one. 
As the   stochastic approach seems to  reproduce the terms with the largest power of the IR logarithms
 at each
order in the coupling constant (leading log approximation) \cite{a1,TW}, it is therefore 
 useful to study the 
late time contributions of the correlation
functions of $\zeta$ 
and to assess whether these can be computed by a classical approximation.

The interaction terms of the comoving curvature perturbation $\zeta$ are purely derivative type interactions, which has for some time hindered a progress in solving the stochastic equations for $\zeta$. In the present paper, we will attempt to make some modest progress on that front. We believe that this can shed some additional light on the IR issues, although a very simple different semi-classical approach to study IR loop corrections to the comoving curvature perturbation and the tensor perturbations has been developed in Ref.~\cite{loops7} (see also \cite{ir6,ir7} for further discussions of this approach).

For small internal momenta, one can derive 
the interaction terms for the various degrees
of freedom in a specific model of inflation, 
and then use the conditions of the theorem of  Ref. \cite{loops1}
to decide whether these interactions can lead to anomalous late time contributions. For single field
inflation (possibly together with  free massless scalar fields $\sigma$'s coupled to it) 
it was  shown in Ref. \cite{loops1} that the
interactions do obey the conditions, and therefore do not give anomalous late time contributions to
all orders. This can be compared with \cite{con}, where  it was shown  that
$\zeta$ is indeed conserved after horizon exit as a consequence of the fact that the system
possesses a time translational invariance in e-folding time when time derivatives
dominate over spatial ones. 
The stochastic approach (which in any case has nothing to say about 
the ultraviolet loops) may offer an alternative way to study the late time behaviour of the correlators
of $\zeta$ and to show that the resummed IR 
loop corrections from these modes that
crossed the horizon arbitrarily in the past do not give rise to any additional time dependence of the
two-point function of the comoving curvature perturbation, 
which therefore stays constant once the mode $k$ is outside of the horizon.

The goal of this paper is   to obtain  the equation governing the probability
$P(\zeta,t)$ that, at a given point, the comoving curvature perturbation acquires a given value
$\zeta$ at a given time $t$, in the presence of self-interactions. Some aspects of this subject were
investigated in a seminal paper  by Bond and Salopek \cite{salopekgrav} and more recently in Refs. \cite{rig,tye,win,tol,stoc1,stoc2}.
As we shall see, the
equation for $P(\zeta,t)$ is  not a diffusive-like equation, that is a Fokker-Planck equation,  but a
generalization of it, the so-called Kramers-Moyal equation. It
 contains an infinite number of derivative with respect to $\zeta$
(the Fokker-Planck equation contains up to two derivatives). This is an unavoidable consequence of the fact that  
self-interactions of the comoving curvature perturbation are present. A key point we would like
to anticipate about the Kramers-Moyal equation is that it will prove to be  convenient to derive 
it starting from  the probability $P(\zeta,\dot{\zeta},t)$, {\it i.e.}  
in terms of two variables, $\zeta$ and  $\dot\zeta$. The stochastic motion
will then be characterized by the fact that 
 $\dot\zeta$   decays rapidly, within a few Hubble times,  to an equilibrium value
due to the presence of a nonvanishing potential $V(\dot{\zeta})$. As we will be  
only interested in times larger than a few Hubble times,  the fast variable
$\dot\zeta$ will be integrated out, leaving us with 
 an effective equation for the probability of the slow variable $\zeta$. This   adiabatic procedure
is  similar to what is done in field theory where heavy degrees of freedom are integrated out resulting in  an effective Lagrangian describing the dynamics of the light fields. The fact that  $\dot\zeta$ rapidly settles to its equilibrium value allows the effective diffusion constant of the $\zeta$ field
to reach a constant (in time) value, thus avoiding an  anomalous  growth of the cumulants of 
$\zeta$ at large times.

The paper is organized as follows. In section 2 we derive the generic, but exact Kramers-Moyal equation 
from the a path-integral formulation. In section 3 we will discuss an alternative derivation of the
Kramers-Moyal equation through the adiabatic elimination of fast variables in the case in which the
self-interaction of the curvature perturbation is dominated by of a  cubic time derivative nature. Section 4
deals with the case in which the curvature perturbation is coupled to an extra scalar degree of freedom.
All these examples are studied under the assumption of neglecting the time evolution of the
slow-roll parameters. In section 
section 5 we relax this assumption and study a particular  case of slow-roll single field inflation.
Finally, section 6 offers our conclusions.

\section{The Kramers-Moyal  equation for  the curvature \\ perturbation}
\noindent
The scope of this section is to describe a generic method to
obtain the  exact equation governing the probability of finding a given value of the
comoving curvature perturbation $\zeta$ at a given time $t$. The method is based on the 
path-integral formulation whose technical details can be found in Refs. \cite{MR1,MR2}.

Before analyzing in detail the case of the comoving curvature perturbation, we will first discuss the more 
standard example of a self-interacting scalar field $\Phi$ with potential $V(\Phi)$ in de Sitter space, {\it i.e.} we assume that the Hubble rate $H$ is constant in time.  
Its field equation is given by

\begin{equation}
\label{a}
\ddot{\Phi}+3H\dot{\Phi}-\frac{1}{a^2}\nabla^2\Phi+V'(\Phi)=0\, .
\end{equation}
The stochastic approach is based on the fact that the IR part of the field may be considered as a classical field satisfying a Langevin-like equation \cite{starob}.
To find the equation for the infrared  part of the field

\begin{equation}
\phi({\bf x},t)=\int\frac{d^3k}{(2\pi)^3}\,\theta\left(Ha-k\right)\left[e^{i{\bf k}\cdot{\bf x}}\phi_{\bf k}(t)a_{\bf k}+
e^{-i{\bf k}\cdot{\bf x}}\phi^*_{\bf k}(t)a^{\dagger}_{\bf k}\right]\, ,
\end{equation}
one can follow the rules rigously obtained Ref. \cite{TW} applied to Eq. (\ref{a}): 

\begin{itemize}
\item  At each order in  the field, retain only the term with the smallest number of derivatives; 

\item for the linear terms in the field, each time derivative has a stochastic source substracted. 
\end{itemize}
This set of rules gives

\begin{equation}
\ddot\phi+3H\dot\phi-\frac{1}{a^2}\nabla^2\phi\rightarrow 3H\dot\phi\rightarrow 3H\left(\dot\phi-\eta_\phi\right)\,,\,\,\,
V'(\phi)\rightarrow V'(\phi)\, ,
\ee
where

\be
\eta_\phi({\bf x},t)=H\,\int\frac{d^3k}{(2\pi)^3}\,k\,\delta_D\left(k-aH\right)\left[e^{i{\bf k}\cdot{\bf x}}\phi_{\bf k}(t)a_{\bf k}+
e^{-i{\bf k}\cdot{\bf x}}\phi^*_{\bf k}(t)a^{\dagger}_{\bf k}\right]\, ,
\ee
is a Gaussian stochastic  noise. Its two-point correlator computed over many
ensemble realizations is 

\be
\langle \eta_\phi({\bf x},t)\eta_\phi({\bf x},t')\rangle=\frac{H^3}{4\pi^2}\delta_D(t-t')\ .
\ee
The behaviour of the 
IR part of the field therefore is described by a Langevin-like equation

\be
\dot\phi({\bf x},t)=\eta_\phi({\bf x},t)-\frac{V'(\phi)}{3H}\, .
\ee
The IR part of the field $\phi$ performs a random walk under the action of the stochastic noise
and, at the same time, it suffers a classical force caused by its self-interaction. Physically, the stochastic  force is caused by the ultraviolet modes that are instantaneously redshifting across the horizon at time $t$. 
It is Markovian  because each increment is uncorrelated with the previous one. This is due to the choice of the
step function in momentum space to select the IR modes. Any other choice of the smoothing function
would lead to a non-Markovian noise and to memory effects in the stochastic motion which will require a proper treatment \cite{MR1}.

In order to obtain the Fokker-Planck equation describing the probability $P(\phi,t)$ that the IR part of the field
acquires a given value $\phi$ at a given time $t$, we exploit the path-integral formulation.
We consider an ensemble of
trajectories all starting at $t=0$ from the  initial position
$\phi({\bf x},0) =\phi_0$ 
and we follow them for a time $t$.  
We discretize the interval $[0,t]$ in steps
$\Delta t=\varepsilon$, so $t_k=k\varepsilon$ with $k=1,\ldots n$, and $t_n\equiv t$. 
A trajectory is  then defined by
the collection of values $\{\phi_1,\ldots ,\phi_n\}$, such that $\phi(t_k)=\phi_k$ (we do not indicate anymore  the spatial variable from now on).
The probability density in the space of  trajectories is 
\be\label{defW}
W(\phi_0;\phi_1,\ldots ,\phi_n;t_n)\equiv \langle
\delta_D(\phi(t_1)-\phi_1)\ldots \delta_D (\phi(t_n)-\phi_n)\rangle\, .
\ee
Our basic object will be
\be\label{defPi}
P (\phi_n,t_n)
 \equiv\int_{-\infty}^{\infty} d\phi_1\ldots \int_{-\infty}^{\infty}d\phi_{n-1}\, 
W(\phi_0;\phi_1,\ldots ,\phi_{n-1},\phi_n;t_n)\, .
\ee
Using the integral representation of the Dirac delta
\be
\delta_D (x)=\int \frac{d\lambda}{2\pi}\, e^{-i\lambda x}\, ,
\ee
we write Eq. (\ref{defW}) as
\be\label{compW}
W(\phi_0;\phi_1,\ldots ,\phi_n;t_n)=\int\frac{d\lambda_1}{2\pi}\ldots\frac{d\lambda_n}{2\pi}\, 
e^{i\sum_{i=1}^n\lambda_i\phi_i} \langle e^{-i\sum_{i=1}^n\lambda_i\phi(t_i)}\rangle
\, .
\ee
We must therefore compute
\be
e^Z\equiv \langle e^{-i\sum_{i=1}^n\lambda_i\phi(t_i)}\rangle \, .
\ee
This is a well-known object both
in quantum field theory and in statistical mechanics, 
since it is  the generating
functional of the connected Green's functions, see {\it e.g.}
\cite{Stratonovich}. To a field theorist
this is even more clear
if we define the ``current'' $J$ from $-i\lambda =\varepsilon J$, and we
use a continuous
notation, so that
\be
e^Z =\langle e^{i\int {\rm d}t\, J(t)\phi(t)}\rangle \, .
\ee 
Therefore
\begin{eqnarray}
Z&=&\sum_{p=1}^{\infty} \frac{(-i)^p}{p!}\, 
 \sum_{i_1=1}^n\ldots \sum_{i_p=1}^n
\lambda_{i_1}\ldots\lambda_{i_p}\, \langle \phi_{i_1}\ldots\phi_{i_p}\rangle_c 
\nn\\
&=&-i\lambda_i\langle\phi_i\rangle_c -\frac{1}{2}\lambda_i\lambda_j\, \langle \phi_i\phi_j\rangle_c\, 
+\frac{(-i)^3}{3!}\,\lambda_i\lambda_j\lambda_k\, \langle \phi_i\phi_j\phi_k\rangle_c \nn\\
&&+\frac{(-i)^4}{4!} \, \lambda_i\lambda_j\lambda_k\lambda_l\,
\langle \phi_i\phi_j\phi_k\phi_l\rangle_c
+\ldots\, ,
\end{eqnarray}
where the sum over $i,j,\ldots$ is understood.
This gives
\be
\label{P}
P(\phi_n,t_n)=\int_{-\infty}^{\infty}\, d\phi_1\ldots \int_{-\infty}^{\infty}\,d\phi_{n-1}\,
\int\,\frac{d\lambda_1}{2\pi}\ldots\frac{d\lambda_n}{2\pi}\, e^{i\lambda_i\phi_i +\sum_{p=1}^{\infty} \frac{(-i)^p}{p!}\, 
\sum_{i_1=1}^n\ldots \sum_{i_p=1}^n
\lambda_{i_1}\ldots\lambda_{i_p}\,
\langle \phi_{i_1}\ldots\phi_{i_p}\rangle_c}\, .
\ee
To get the equation that governs the evolution of the probability $P(\phi_n,t_n)$, we take the time derivative of Eq. (\ref{P})

\begin{eqnarray}
\label{deriFP1}
\frac{\partial P}{\partial t_n}&=&
- i
\frac{\partial \langle\phi_k\rangle_c}{\partial t_n}
\int_{-\infty}^{\infty} d\phi_1\ldots d\phi_{n-1}\,
\int\,\frac{d\lambda_1}{2\pi}\ldots\frac{d\lambda_n}{2\pi}\
\lambda_k
 e^{i\lambda_i\phi_i +\sum_{p=1}^{\infty} \frac{(-i)^p}{p!}
\sum_{i_1=1}^n\ldots \sum_{i_p=1}^n
\lambda_{i_1}\ldots\lambda_{i_p}
\langle \phi_{i_1}\ldots\phi_{i_p}\rangle_c}\, \nonumber\\
&-&
\frac{1}{2} 
\frac{\partial \langle\phi_k\phi_l\rangle_c}{\partial t_n}\hspace{-0.1cm}
\int_{-\infty}^{\infty}\hspace{-0.1cm} d\phi_1\cdots d\phi_{n-1}\hspace{-0.1cm}
\int\frac{d\lambda_1}{2\pi}\cdots\frac{d\lambda_n}{2\pi}\
\lambda_k\lambda_l
 e^{i\lambda_i\phi_i +\sum_{p=1}^{\infty} \frac{(-i)^p}{p!} 
\sum_{i_1=1}^n\ldots \sum_{i_p=1}^n
\lambda_{i_1}\ldots\lambda_{i_p}
\langle \phi_{i_1}\ldots\phi_{i_p}\rangle_c}\, \nonumber\\
&+&\cdots.
\end{eqnarray}
Using the replacement $\lambda_i\exp\{i\lambda_i \phi_i\}=-i\partial_{\phi_i}\exp\{i\lambda_i \phi_i\}$ (the index $i$ is not summed over), 
inside the integrals we can replace
$\lambda_i\ra  -i\partial_{\phi_i}$.
Since  we integrate over
$d\phi_1, \ldots d\phi_{n-1}$, but not over $d\phi_n$,
if the generic index $i\neq n$ the term
$\pa_{\phi_i}$, when integrated over $d\phi_i$, is a total derivative and gives
zero, because at the boundaries  $\phi_i=\pm\infty$ the integrand vanishes
exponentially, and the only contribution comes from $i=n$.
The equation (\ref{deriFP1}) reduces therefore to (setting $t_n=t$ and $\phi_n=\phi$)

\be\label{P1}
\frac{\pa P(\phi,t)}{\pa t}
=\sum_{p=1}^{\infty}\frac{(-1)^p}{p!}\, \frac{\partial{\mu}_p(t)}{\partial t} \,
\frac{\pa^p P(\phi,t)}{\pa \phi^p}\, ,
\ee
where

\be
\mu_p(t)=\langle \phi^p(t)\rangle_c
\ee
is the $p$-th connected cumulant.
Equation (\ref{P1}) is called the Kramers-Moyal (KM) equation or ``the stochastic
equation'', and is well known in the theory of stochastic
processes \cite{Stratonovich,Risken}. Its solution requires the exact knowledge of the cumulants and in this sense the KM equation does not contain more information than the generating functional.
Furthermore,  
  Pawula's theorem \cite{pawula}  states that if we assume 
$\mu_p=0$ for some even $p$, then we are actually assuming $\mu_p$ to be zero for all
$p\geq 3$. Thus,  it is logically inconsistent to retain more than two terms in the KM
expansion unless all of the terms are retained. Nevertheless 
inconsistencies may  lead to tiny numerical errors and one usually reduces the KM equation (\ref{P1})
to a Fokker-Planck equation

\be\label{P2}
\frac{\pa P(\phi,t)}{\pa t}
=\frac{1}{3H}\frac{\pa}{\pa\phi}\left[V'(\phi)P(\phi,t)\right]+\frac{H^3}{8\pi^2}\frac{\pa^2}{\pa\phi^2}P(\phi,t)\, ,
\ee
where one has exploited the fact that  
at leading order in the coupling constant(s) the connected two-point correlator is dominated by the 
Gaussian and Markovian stochastic noise.
At small times the diffusive term dominates 

\be\label{P3}
\frac{\pa P(\phi,t)}{\pa t}
\simeq \frac{H^3}{8\pi^2}\frac{\pa^2}{\pa\phi^2}P(\phi,t)\, 
\ee
and the scalar field performs a diffusive random walk  such that $\langle\phi^2\rangle\sim H^3 t$. At larger times, the friction term intervenes
and the  equation might have a  non-perturbative stationary late time solution induced by the fact that the scalar force eventually balances the tendency of inflationary particle production to force the scalar
up its potential \cite{sy}

\be
\label{poi}
P_{\rm st}(\phi,t\gg H)\sim {\rm exp}\left(-8\pi^2V(\phi)/3H^4\right)\, .
\ee
To estimate the time scale at which the stationary solution is reached one has to specify the
potential. For instance, for a quartic potential $V(\phi)=\lambda\phi^4$, the 
variance of the field is given by  $\langle\phi^2\rangle\sim H^2/\sqrt{\lambda}$. This means that it takes
a time $t\sim H^{-1}/\sqrt{\lambda}$ for the stationary regime to be reached.
 Notice that if the
field $\Phi$ is the inflaton, the   eternal inflation regime is achieved when the potential is flat enough that quantum corrections always
dominate over the classical motion. In such a case,    the diffusive term
dominates and the variance of the field grows linearly in time. 

The derivation of the equation governing the probability $P(\zeta,t)$ through the path-integral formulation follows basically the same steps described for the self-interacting scalar
field in de Sitter space. In order to find a simple solvable toy model, we find it convenient
to adopt the effective field  theory of inflation \cite{eff}
which makes it possible to explore in full generality all the possible self-interactions of the inflaton, and in particular to clearly see that the inflaton can have large self-interactions without spoiling the background quasi de Sitter solution. It was shown that the self-interactions can be parametrically larger than the one mediated by gravity and therefore we can in some cases concentrate on those without worrying about metric fluctuations. Note, however, that in doing this, we are exactly neglecting the interactions that leads to large IR effects. The interactions that leads to large IR effects are subdominant in the slow-roll expansion, and vanishes in the decoupling limit of the effective approach. So, while we don't expect to see any large IR effects in this limit, it is still useful for providing a simple solvable toy model for testing the stochastic approach. 

The effective Lagrangian is a function of the field $\pi$ which represents the Goldstone boson of time translations which are spontaneously broken during inflation. It is related to the $\zeta$ metric fluctuation at linear level by the simple relation $
\zeta=-H\pi$. The general derivation of the Lagrangian for $\pi$  can be found in Ref. \cite{eff} and we do not present it here. It suffice to say that, for instance,  at the cubic level there are in general two interactions of comparable strength, $\dot\pi^3$ and $\dot\pi(\nabla \pi)^2$, each one accompanied by quartic interactions due to symmetry reasons. Upon tuning, one  can make the term containing $\dot\pi^3$ parametrically larger (smaller) than the one containing $\dot\pi(\nabla \pi)^2$. Following the notation of \cite{eff}, the action for $\zeta$ reduces to  the form (see also \cite{chen})

\be
\label{eq:action}
S=\int d^4x\; a^3\left[-\frac{\dot H}{c_s^2 H^2}\mpl^2\left(\dot\zeta^2-\frac{c_s^2}{a^2}(\nabla\zeta)^2\right)-
\frac{\dot H\mpl^2}{H^3}\left(1-\frac{1}{c_s^2}\right)\left(\dot\zeta^3-\dot\zeta\frac{(\nabla\zeta)^2}{a^2}\right)+\frac{4}{3}\frac{M_3^4}{H^3}\dot\zeta^3+\cdots\right]\ ,
\ee
where $c_s$ is the sound speed, $\dot H=-\epsilon H^2$ defines the slow-roll parameter $\epsilon$,  $M_3$ is a mass term and the dots stand for the terms which are 
proportional to the slow-roll parameters and to the higher-order terms in perturbation theory. The IR part of the field can be defined as usual

\begin{equation}
\zeta({\bf x},t)=\int\frac{d^3k}{(2\pi)^3}\,\theta\left(Ha-k\right)\left[e^{i{\bf k}\cdot{\bf x}}\zeta_{\bf k}(t)a_{\bf k}+
e^{-i{\bf k}\cdot{\bf x}}\zeta^*_{\bf k}(t)a^{\dagger}_{\bf k}\right]
\end{equation}
It is subject to a stochastic noise

\be
\eta_\zeta({\bf x},t)=H\,\int\frac{d^3k}{(2\pi)^3}\,k\,\delta_D\left(k-aH\right)\left[e^{i{\bf k}\cdot{\bf x}}\zeta_{\bf k}(t)a_{\bf k}+
e^{-i{\bf k}\cdot{\bf x}}\zeta^*_{\bf k}(t)a^{\dagger}_{\bf k}\right]\, ,
\ee
whose two-point correlator computed over many
ensemble realizations is computed using linear perturbation theory and is defined  as  

\be
\label{ml}
\langle \eta_\zeta({\bf x},t)\eta_\zeta({\bf x},t')\rangle=\frac{D_\zeta}{9H^2}\delta_D(t-t')\ ,
\ee
where
\be
D_\zeta=\frac{9H^5}{8c_s\epsilon\pi^2\mpl^2}\, .
\ee
The corresponding motion is stochastic and the  probability that the curvature perturbation
$\zeta$ acquires the value $\zeta_n$ after $n$-steps at time $t_n$ is given by

\be
\label{Pzeta}
P(\zeta_n,t_n)=\int_{-\infty}^{\infty}\, d\zeta_1\ldots \int_{-\infty}^{\infty}\,d\zeta_{n-1}\,
\int\,\frac{d\lambda_1}{2\pi}\ldots\frac{d\lambda_n}{2\pi}\, e^{i\lambda_i\zeta_i +\sum_{p=2}^{\infty} \frac{(-i)^p}{p!}\, 
\sum_{i_1=1}^n\ldots \sum_{i_p=1}^n
\lambda_{i_1}\ldots\lambda_{i_p}\,
\langle \zeta_{i_1}\ldots\zeta_{i_p}\rangle_c}\, .
\ee
Taking the derivative with respect to $t_n$ and setting at the end $t_n=t$ and $\zeta_n=\zeta$ we obtain the exact Kramers-Moyal
 equation for the comoving curvature perturbation

\be\label{Pzeta1}
\frac{\pa P(\zeta,t)}{\pa t}
=\sum_{p=2}^{\infty}\frac{(-1)^p}{p!}\, \frac{\partial{\mu}_p(t)}{\partial t} \,
\frac{\pa^p P(\zeta,t)}{\pa \zeta^p}\, ,
\ee
where

\be
\mu_p(t)=\langle \zeta^p(t)\rangle_c
\ee
is the $p$-th connected cumulant of the perturbation $\zeta$. 
The sum over the index $p$
is restricted to $p\geq 2$, the term proportional to one derivative with respect to $\zeta$ is not present.
This is because we require that  $\zeta$, being a perturbation, has an expectation value which vanishes  at any order in perturbation theory. If we start with the initial condition $\langle\zeta\rangle_c=0$ at $t=0$, then equation (\ref{Pzeta1}) insures that it remains zero at all times

\be
\frac{\partial}{\partial t}\,\langle\zeta\rangle_c=\int\,d\zeta\,\zeta\,\frac{\partial}{\partial t}\,P(\zeta,t)=0\, ,
\ee
as one can  see by  plugging in expression (\ref{Pzeta1}) and integrating by parts.
The  KM equation (\ref{Pzeta1}) for the curvature perturbation does not contain any more information than the full generating functional of $\zeta$.  In this sense, there is complete analogy with what
happens in field theory where the  knowledge of the full dynamics of a system requires
the knowledge of all the $n$-point connected correlators. The KM equation describes  the  time-flow of the functional  in the same way the renormalization group equations describe the flow of the
generating functional and the $n$-point correlators with energy. As in field theory where one proceeds
performing a perturbation expansion in some coupling constants, 
 one can try to  perturbatively simplify the KM equation to solve for the probability. 
Even if  Pawula's theorem \cite{pawula} implies
that throwing away terms in Eq. (\ref{Pzeta1}) with a number of derivatives with respect to
$\zeta$ higher than two to obtain a diffusive Fokker-Planck equation is logically  inconsistent, it might  lead to tiny errors \cite{rv}, especially if one considers the case of the evolution during the last 60 e-folds
of inflation.  For a cubic interaction one could make an expansion in the coupling 
constant(s) keeping only the cubic term $p=3$ and approximating the 
two-point cumulant to its tree-level value deduced from  (\ref{ml}),

\be
\langle \zeta^2\rangle_c=\int_0^t {\rm d}t'\, \frac{D_\zeta}{9 H^2}\, .
\ee 
Alternatively, one can stop at the second derivative with respect to $\zeta$ and obtain a Fokker-Planck equation. Again, as in field theory the perturbative renormalization group evolution of the couplings constants is able to catch and resum the leading logs, the KM equation may be 
able to provide non-perturbative  solutions,  like the stationary one in Eq. (\ref{poi}).
As far as the  initial condition is concerned, Eq. (\ref{Pzeta1}) should be solved with the 
assumption that at $t=0$

\be
\label{con1}
P(\zeta,0)=\delta_D(\zeta)\, .
\ee
Let us pause
for a moment and discuss  the interpretation of such initial condition  \cite{salopekgrav,ir3}.
If we consider only the physical observables within a comoving region
coinciding with our present horizon volume, where several observations indicate
that this patch is highly homogeneous and isotropic, 
 we have to 
 select those
initial conditions  which respect these observations, and therefore select the initial time $t=t_{60}$
to be the time when one is left with the last 60 e-folds or
so of inflation. 
If, on the other side, we  insist in adopting a superlarge box containing many smaller patches
of size similar to our observable  universe, as commented in the introduction, the initial time $t=0$ should be 
taken to be the beginning of inflation.
If so, 
it should be kept in mind that the location of our box
may be untypical and one should quantify how likely it is that the correlators
averaged over the superlarge box coincide with the correlators in
our observable universe.
One should split the computation of the probability in two steps. 
First, the distribution of the values of the
curvature perturbation  inside the superlarge box should be estimated.
This will allow one to compute the probability
$P(\zeta_{60}\left.\right|t_{60})$
that the starting initial condition for the background field is given by
$\zeta_{60}$ in our local observable patch at time
$t_{60}$
when there remain 60 or so e-folds till the end of inflation in that patch.
The distribution of the comoving curvature perturbation in the superlage box, due to the IR
divergences, will be far from being Gaussian. 
However, we do expect that in our local patch initial conditions 
are relatively uniform. From the point of view of our local universe
the IR properties of the superlarge box only enter to provide the probability 
$P(\zeta_{60}|t_{60})$ that 
the value $\zeta_{60}$ is achieved in our patch. At later times the
field value at a given point has a distribution given by
\begin{equation}
	P(\zeta,t)=\int d{\zeta}_{60}\,
	P\left({\zeta}\left.\right | t; t_{60},\zeta_{60}
	\right) P(\zeta_{60}, t_{60}),
\end{equation}
Furthermore, if the comoving curvature  is homogeneously distributed
in our local patch, then $
P\left(\zeta\left.\right | t_{60}; 
	t_{60},\zeta_{60}
	\right)=\delta_D\left(\zeta-\zeta_{60}\right)$ with $\zeta_{60}\simeq 0$. 
The probability $P(\zeta,t)$ 
is generically  highly non-Gaussian, because the  initial conditions are not
Gaussian distributed as large IR corrections are present in the superlarge
box.

\section{An alternative derivation of the Kramers-Moyal equation: the adiabatic elimination of fast variables}
\noindent
The scope of this section is to  offer  an alternative and perturbative derivation of the
KM equation for the curvature perturbation. We will restrict ourselves
to  the case in which large non-Gaussianities are present, {\it i.e.} we will disregard  the self-interaction terms in the Lagrangian proportional to the
slow-roll parameters. The same assumption allows us to utilize the linear relation between $\zeta$ and $\pi$. Again, we note that we don't expect any large IR effects in this limit, since the interactions leading to large IR effects are slow-roll suppressed, but it may nevertheless be a useful limit for testing the methods.

We are  interested in the
behaviour of the long wavelength part of the comoving curvature perturbation. Following the
procedure described in the previous section, we consider the
IR part of $\zeta$ as a classical space-dependent stochastic field which satisfies
a local Langevin-like equation. Furthermore, 
we  suppose that at the cubic level the
interaction term is dominated by the $\dot\zeta^3$ term, {\it i.e.} we suppose that $M^4_3\gg\epsilon H^2\mpl^2$. Following the rules dictated by \cite{TW}, the IR part of the comoving curvature perturbation
satisfies the equation

\be
\label{aa}
\ddot\zeta+3H\dot\zeta-6 c_s^2\frac{M_3^4}{\dot H \mpl^2}\dot\zeta^2=3H \eta_\zeta\, ,
\ee
where $\eta_\zeta$ is the stochastic Markovian noise whose two-point  correlator is

\be
\langle \eta_\zeta({\bf x},t)\eta_\zeta({\bf x},t')\rangle=\frac{H^3}{8c_s\epsilon\pi^2\mpl^2}\delta_D(t-t')\ .
\ee
To determine the equation
governing the probability, it is useful to  
 rewrite Eq. (\ref{aa}) doubling the degrees of freedom \cite{Risken}

\be
\label{e1}
\dot\zeta=v_\zeta\, ,
\ee
\be
\label{e2}
\dot{v}_\zeta+3H v_\zeta+6 c_s^2\frac{M_3^4}{\epsilon H^2 \mpl^2}v_\zeta^2=3H \eta_\zeta\, .
\ee
Following the steps described in the previous section, we may write the KM equation for the
probability 
$P(\zeta,v_\zeta,t)$ as

\be
\label{q}
\frac{\partial P }{\partial t}=-\frac{\partial }{\partial \zeta}\left(v_\zeta P\right)+
\frac{\partial}{\partial v_\zeta}\left[V'(v_\zeta)P\right]+\frac{1}{2}D_\zeta
\frac{\partial^2}{\partial v_\zeta^2}P\, ,
\ee
where 
\be
\label{pot}
V(v_\zeta)=\frac{3}{2} H v_\zeta^2+2 c_s^2\frac{M_3^4}{\epsilon H^2 \mpl^2}v_\zeta^3\, .
\ee
Notice that in the first and second terms of the right-hand side of Eq. (\ref{q}) we might replace $v_\zeta$ by $\left(v_\zeta-\langle v_\zeta\rangle\right)$ and $V'(v_\zeta)$ by $\left(V'(\zeta)-\langle V'(v_\zeta)\rangle\right)$
to account for the fact that 
the comoving curvature perturbation $\zeta$ and $v_\zeta$  must  have vanishing expectation value at all times
at all orders. We will come back to this point later.
The initial condition of Eq. (\ref{q}) is given by

\be
\label{con}
P(\zeta,v_\zeta,0)=\delta_D(\zeta)\delta_D(v_\zeta)\, .
\ee
In order to solve Eq. (\ref{q}) we may proceed using the method of the adiabatic elimination of the
fast variables \cite{kaneko}. The idea is that the variable $v_\zeta$ will decay rapidly to an equilibrium value
due to the presence of a nonvanishing potential $V(v_\zeta)$. Inspecting the energy scales
in the potential, it is easy to get convinced that the equilibrium situation should be reached
after one Hubble time, $t_{\rm eq}\sim H^{-1}$. 
As we are only interested in times larger than the decay time $t_{\rm eq}$ of the fast variable
$v_\zeta$, the process described by Eqs. (\ref{e1}) and (\ref{e2}) is then mainly described by the motion of the slow variable $\zeta$. Hence we may say that the slow variable slaves the fast variable \cite{haken}. The slow variable is then the  important or relevant variable, since the fast variable becomes irrelevant for the above time scale because it can be expressed by the slow variable. Integrating out 
the fast variable $v_\zeta$ will allow us to get an effective equation for the probability of the slow variable $\zeta$ with a procedure similar to what is done in field theory where heavy degrees of freedom are integrated out resulting in  an effective Lagrangian describing the dynamics of the light fields.

We first look at the operator

\be
L_{v_\zeta}=\frac{\partial}{\partial v_\zeta}V'(v_\zeta)+\frac{1}{2}D_\zeta
\frac{\partial^2}{\partial v_\zeta^2}\, .
\ee
We introduce the following operators

\begin{eqnarray}
a&=&\sqrt{\frac{D_\zeta}{2}}\frac{\partial}{\partial v_\zeta}+\sqrt{\frac{1}{2 D_\zeta}}V'\,,\nonumber\\
\hat{a}&=-&\sqrt{\frac{D_\zeta}{2}}\frac{\partial}{\partial v_\zeta}+\sqrt{\frac{1}{2D_\zeta}}V'\,,
\end{eqnarray}
which, for natural boundary conditions are the adjoints of each other, {\it i.e.}, $\hat{a}=a^\dagger$. The operator $L_{v_\zeta}$ 
 can be recast in the form

\be
L_{v_\zeta}=-{\rm exp}\left(-V/D_\zeta\right)\,a^\dagger a\,{\rm exp}\left(V/D_\zeta\right)\, .
\ee
If we denote by $\psi_n(v_\zeta)$  the eigenfunctions of the Hermitian operator $a^\dagger\,a$

\be
a^\dagger a\, \psi_n=\lambda_n\,\psi_n\, ,
\ee
then all $\lambda_n$ must be real and non-negative. Furthermore 
$\Psi_n={\rm exp}\left(-V/D_\zeta\right)\psi_n$ are  the eigenfunctions of the operator $L_{v_\zeta}$ with  eigenvalues $-\lambda_n\leq 0$.  
For $n=0$ we have the stationary solution induced by the zero mode

\be
\Psi_0(v_\zeta)={\cal N}_0\left(-2V/D_\zeta\right)\, ,\,\,\,\lambda_0=0\, ,
\ee
where ${\cal N}_0$ is the normalization factor. 
The eigenfunctions $\Psi_n=\sqrt{\Psi_0}\psi_0$ and the eigenfunctions $\Psi_n^\dagger=
\psi_n/\sqrt{\Psi_0}=
\Psi_n/\Psi_0$ of the adjoint operator $L_{v_\zeta}^\dagger$  form an orthonormal basis satisfying

\be
\label{ort}
\int_{-\infty}^{\infty}\, \Psi_n^\dagger(v_\zeta)\Psi_m(v_\zeta)\,dv_\zeta=\delta_{mn}\, ,
\ee
and
\be
\sum_{n=0}^{\infty}\Psi_n^\dagger(v_\zeta)\Psi_m(v'_\zeta)=\delta_D\left(v_\zeta-v'_\zeta\right)\, .
\ee
Notice in particular that $\Psi_0^\dagger(v_\zeta)=1$. 
We now expand the distribution function $P(\zeta,v_\zeta,t)$ into the  complete set $\Psi_n$ of the operator $L_{v_\zeta}$

\be 
P(\zeta,v_\zeta,t)=\sum_{m=0}^{\infty}\,P_m(\zeta,t)\Psi_m(v_\zeta)\, .
\ee
We insert this expansion into the KM equation (\ref{q}) multiplying the resulting equation by $\Psi_n^\dagger$ and then integrating over the variable $v_\zeta$, thus obtaining

\be
\label{r}
\left(\frac{\partial}{\partial t}+\lambda_n\right)P_n=\sum_{m=0}^\infty \, L_{n,m}\,P_m
\ee
with

\be
L_{n,m}= \int_{-\infty}^\infty\,\Psi_n^\dagger(v_\zeta)\, L_\zeta\,\Psi_m(v_\zeta)\,dv_\zeta\, ,
\ee
and

\be
L_\zeta=-\frac{\partial }{\partial \zeta}v_\zeta \, .
\ee
The infinite set of equations (\ref{r}) is exact and, as expected, it contains an infinite number of derivatives with respect to $\zeta$ once an iterative procedure is adopted to solve for the $P_m$'s. At times $t\gg t_{\rm eq}\sim H^{-1}$, we may use the following approximation procedure. Knowing that the eigenvalues $\lambda_n$ are positive for all $n\geq 1$, we  may neglect in (\ref{r}) the time derivative in the equations
for $n$ larger than zero, that is

\begin{eqnarray}
\frac{\partial}{\partial t}{P}_0&=&\sum_{m=0}^\infty \, L_{0,m}\,P_m=L_{0,0}P_0+\sum_{m=1}^\infty \, L_{0,m}\,P_m\, ,\nonumber\\
P_n&=&\lambda_n^{-1}\sum_{m=0}^\infty \, L_{n,m}\,P_m
\nonumber\\
&=&\lambda_n^{-1}\,L_{n,0}\,P_0+\lambda_n^{-1}\sum_{m=1}^\infty \, L_{n,m}\,P_m\,\nonumber\\
&\simeq&\lambda_n^{-1}\,L_{n,0}\,P_0+\lambda_n^{-1}\lambda_m^{-1}\sum_{m=1}^\infty \, L_{n,m}\,L_{m,0}P_0+\cdots\, ,
\label{mm}
\end{eqnarray}
where the last passage is possibile because the $P_m$ for $m\geq 1$ decays in time faster than
$P_0$ for which $\lambda_0=0$. We thus obtain that the equation governing the probability 

\begin{eqnarray}
P(\zeta,t)&=&\int_{-\infty}^{\infty}\,dv_\zeta\, P(\zeta,v_\zeta,t)=\sum_{n=0}^{\infty}\int_{-\infty}^{\infty}\,dv_\zeta \Psi_n(v_\zeta)\,P_n(\zeta,t)\nonumber\\
&=&\sum_{n=0}^{\infty}\int_{-\infty}^{\infty}\,dv_\zeta \Psi^\dagger_0(v_\zeta)\,\Psi_n(v_\zeta)\,P_n(\zeta,t)\nonumber\\
&=&P_0(\zeta,t)\, ,
\end{eqnarray}
is the KM  equation
\begin{eqnarray}
\label{ppp}
\frac{\partial}{\partial t}P(\zeta,t)&=&L_{\rm KM}\,P(\zeta,t)\nonumber\\
&=&\sum_{n=1}^\infty\,\lambda_n^{-1}\,L_{0,n}
\left(L_{n,0}+\lambda_m^{-1}\sum_{m=1}^\infty \, L_{n,m}L_{m,0}+\cdots\right)
\,P(\zeta,t)\,\nonumber\\
&=&\frac{1}{2}\,\frac{D_\zeta^{\rm eff}}{9H^2}\frac{\partial^2}{\partial \zeta^2}P(\zeta,t)+\frac{1}{3!}\,S^{\rm eff}_\zeta\frac{\partial^3}{\partial \zeta^3}P(\zeta,t)+\cdots\, ,
\end{eqnarray}
which contains an infinite number of derivatives with respect to $\zeta$. 
The first two coefficients of the KM  expansion are

\be
\label{gg}
\frac{1}{2}\,\frac{D_\zeta^{\rm eff}}{9H^2}=
 \sum_{n=1}^\infty\,\lambda_n^{-1}\left(\int_{-\infty}^{\infty}\, v_\zeta\,\Psi_n(v_\zeta)\,dv_\zeta\right)
\left(\int_{-\infty}^{\infty}\, \Psi_n^\dagger\, v_\zeta\,\Psi_0(v_\zeta)\,dv_\zeta\right)\, ,
\ee
and
\be
\frac{1}{3!}\,S_{\zeta}^{\rm eff}=
 \sum_{n=1}^\infty\lambda_n^{-1}\left(\int_{-\infty}^{\infty} v_\zeta\Psi_n(v_\zeta)\,dv_\zeta\right)
\sum_{m=1}^\infty\lambda_m^{-1}\left(
\int_{-\infty}^\infty\Psi_n^\dagger(v_\zeta)\, v_\zeta\,\Psi_m(v_\zeta)\,dv_\zeta\right)
\left(\int_{-\infty}^{\infty}\Psi_m^\dagger v_\zeta\Psi_0(v_\zeta)\,dv_\zeta\right)\, .
\ee
Notice that the term proportional to the
first derivative with respect to $\zeta$ coming from  $L_{0,0}$ vanishes which is justified by requiring that at all orders in perturbation theory the expectation value of $\zeta$ must vanish. Indeed, we should replace the operator $L_\zeta=v_\zeta\partial/\partial\zeta$ with $(v_\zeta-\langle v_\zeta\rangle)\partial/\partial\zeta$. This implies that $L_{0,0}$ should be replaced by

\begin{eqnarray}
\hat{L}_{0,0}&=&\left(\int_{-\infty}^\infty d v_\zeta\Psi_0^\dagger v_\zeta \Psi_0-\int_{-\infty}^\infty d\zeta'\int_{-\infty}^\infty dv_\zeta \sum_{n=0}^\infty P_n(\zeta') v_\zeta\Psi_n\right)\frac{\partial}{\partial\zeta}\nonumber\\
&=&\left(\int_{-\infty}^\infty d v_\zeta\Psi_0^\dagger v_\zeta \Psi_0-\int_{-\infty}^\infty d\zeta' P_0(\zeta')\int_{-\infty}^\infty dv_\zeta\Psi_0^\dagger v_\zeta\Psi_0-\int_{-\infty}^\infty d\zeta' P_1(\zeta')\int_{-\infty}^\infty dv_\zeta\Psi_0^\dagger v_\zeta\Psi_1+\cdots\right)\frac{\partial}{\partial\zeta}\nonumber\\
&=&0\, ,
\end{eqnarray}
where we have used the fact that  $\Psi_0^\dagger=1$ and  $P_0(\zeta,t)=P(\zeta,t)$ is normalized to unity, and that the $P_n$'s with $n\geq 1$, according to (\ref{mm}), are expressed in terms of derivatives of $P_0$ with respect to $\zeta$ and therefore their total integrals vanish. 
Due to the
Pawula's theorem \cite{pawula}, we are not allowed to disregard the terms with a number of derivatives
with respect to $\zeta$ higher or equal to three in the KM equation. Nevertheless, if we are only interested in the
time behaviour of the variance $\langle\zeta^2\rangle$ 

\be
\frac{\partial}{\partial t}\langle\zeta^2\rangle=\int \,d\zeta\,\zeta^2\,\frac{\partial}{\partial t} P(\zeta,t)\, ,
\ee
we can restrict ourselves to the term with two derivatives with respect to $\zeta$. This gives

\be
\langle\zeta^2\rangle = \frac{D_\zeta^{\rm eff}}{9H^2}t\, .
\ee
The linear dependence originates by our  initial assumption that the slow-roll parameters are tiny so that 
their time evolution is negligible; 
otherwise we should use as time variable not $t$ but $N=\ln\,a$ 
\cite{starob} and get

\be
\langle\zeta^2\rangle = \int^N dN'\,
\frac{H^2}{16c_s\epsilon\pi^2\mpl^2}\, .
\ee
The coefficient $D_\zeta^{\rm eff}$ in (\ref{gg}) gets contribution only at second order in the perturbative parameter
$\lambda=(2 c_s^2M_3^4/\epsilon H^2 \mpl^2)$ in front of the cubic term in the potential
(\ref{pot}). Indeed, the ${\cal O}(\lambda)$ correction to the eigenvalues  $\lambda_n$ vanishes,

\be
\delta\lambda_n=\lambda\langle\Psi^{(0)\dagger}_n|v_\zeta^3|\Psi^{(0)}_n\rangle=0\, ,
\ee
 where
$\Psi_n^{(0)}$ are the eigenfunction of the linear problem (see below); the correction 
to the eigenfunctions are given by

\begin{eqnarray}
\label{series}
\Psi^{(1)}_n&=&\Psi_n^{(0)}+{\cal O}(\lambda)\left[
\sqrt{n(n-1)(n-2)}\Psi_{n-3}^{(0)}\right.\nonumber\\
&+&\left. 9n^{3/2}\Psi_{n-1}^{(0)}-9(n+1)^{3/2}\Psi_{n+1}^{(0)}-\sqrt{(n+1)(n+2)(n+3)}\Psi_{n+3}^{(0)}\right]\, .
\end{eqnarray}
Inspecting the expression (\ref{gg}), we can first insert the
expression (\ref{series}) in the 
first integral in the right hand side while anywhere else we pick up the eigenfunctions at the zero-th order.
This  selects an odd $n$ from the second integral since $v_\zeta$ is odd and $\Psi_0^{(0)}$ is even. 
However, from (\ref{series}) we read that the corrections to the  the eigenfunctions with $n$ odd  receive corrections  only from even   functions $\Psi_{p}^{(0)}$ and therefore this correction vanishes.
We now perturb the second integral in the expression (\ref{gg}). Being $v_\zeta$ proportional to
$\Psi_1^{(0)\dagger}$ (see below)  this selects $n=1$ in the first integral; again the corrections in $\Psi_1^{(1)}$ are all even and this makes the second integral vanishing. The same happens if  we perturb $\Psi_0$ in the second integral as it gets corrections only from odd functions. We conclude that the effective diffusion coefficient receives corrections only at second order in the coupling constants.
On the other hand it is easy to see that $S_\zeta^{\rm eff}$ receives contributions which
are ${\cal O}(\lambda)$. 
This is what we expect 
from computing diagramatically the one-loop correction to the connected two-point cumulants
$\langle\zeta^2\rangle$ and $\langle\zeta^3\rangle$ . The corrections to the diffusion coefficient are time-independent.
 This is intimately linked to the fact that 
the reduction to the KM equation (\ref{ppp})
has been made possible because the fast variable $v_\zeta$ reaches rapidly
a stationary solution and can be integrated out. This conclusion is in fact valid  at any order in perturbation theory
when  the dominant contribution to the non-linearities at any given order $p$ is dominated by
the time-derivative interaction $v_\zeta^p=\dot{\zeta}^p$. In such a case, the potential for the
fast variable $v_\zeta$ is of the form 

\be
V(v_\zeta)=\frac{3}{2}H v_\zeta^2+\sum_{m=1}^p \frac{c_m}{m}v_\zeta^m\, ,
\ee
and the elimination of the fast variable $v_\zeta$ applies. This allows us to conclude
that comoving curvature perturbation performs a diffusive random motion where $\langle \zeta^2\rangle\sim t$ (neglecting the time variation of the slow-roll parameters) even in the presence of time-derivative interactions: after a time scale of the order of the
inverse Hubble time the diffusive coefficient $D_\zeta^{\rm eff}$ saturates to a constant.
In the opposite case in which the $\dot{\zeta}\left(\nabla\zeta\right)^2$ cubic 
term dominates over the $\dot{\zeta}^3$ term, one expects an equation for the probability which is very similar to the Kardar-Parisi-Zhang equation \cite{kpz} encountered in investigating scaling properties and general morphology of
nonequilibrium systems. One possibility of dealing with such a case is of course to compute directly the cumulants and construct the KM equation as described in section 2. A procedure
similar to the one described in this section is technically more demanding and we leave it for future work.

\subsection{A useful check: the linear case}
\noindent
Before concluding this section, it might be interesting to   assess the  validity of the adiabatic procedure followed so far. To do so, we  reduce ourselves to the
 linear case, {\it i.e.} when the potential $V(v_\zeta)$ is quadratic.  Eq. (\ref{q}) with the boundary condition (\ref{con}) is exactly solvable   \

\be
P_{\rm lin}(\zeta,v_\zeta,t)=\frac{1}{2\pi}\frac{1}{\sqrt{{\rm Det}\,M}}{\rm exp}\left(-1/2{ M}^{-1}_{\zeta\zeta}\zeta^2-{M}^{-1}_{\zeta v_\zeta}\zeta v_\zeta-1/2{ M}^{-1}_{v_\zeta v_\zeta}v_\zeta^2\right)\, ,
\ee
where
\begin{eqnarray}
{ M}_{\zeta\zeta}&=&\frac{1}{54}\frac{D_\zeta}{H^2}t\left(6Ht-3+4 e^{-3Ht}-e^{-6Ht}\right)\, ,\nonumber\\
{ M}_{\zeta v_\zeta}&=&\frac{1}{18}\frac{D_\zeta}{H^2}\left(1-e^{-3Ht}\right)^2\, ,\nonumber\\
{ M}_{v_\zeta v_\zeta}&=&\frac{1}{6}\frac{D_\zeta}{H}\left(1-e^{-6Ht}\right)\, .
\end{eqnarray}
It is clear from this expression that the typical time scale by which the variable $v_\zeta$ reaches
the equilibrium value is the inverse of the Hubble time. 
At  large times, $t\gg H^{-1}$, the probability reduces to 

\be
P_{\rm lin}(\zeta,v_\zeta,t)=\frac{1}{2\pi}\left(\frac{27 H^3}{2 D^2_\zeta t}\right)^{1/2}{\rm exp}\left[-(9 H^2/D_\zeta t) \zeta^2\right]{\rm exp}\left[(3H/ D_\zeta t)
\zeta v_\zeta\right]
{\rm exp}\left[-(6 H/ D_\zeta)v_\zeta^2\right]\, .
\ee
Integrating this expression in $v_\zeta$ one finds that at large times

\be
P_{\rm lin}(\zeta,t)=\frac{3}{2\sqrt{2 \pi D_\zeta t}}\,{\rm exp}\left[-(9 H^2/2 D_\zeta t)\zeta^2\right]\, 
\ee
and

\be
\label{o}
\langle\zeta^2\rangle =\int\,d\zeta\,\zeta^2\,P_{\rm lin}(\zeta,t)=\frac{D_\zeta}{9 H^2}t\, .
\ee
Let us now check  that  the same result is obtained using the 
 elimination of the fast variable
$v_\zeta$. In the linear case the equation for $v_\zeta$ reduces to that of the harmonic oscillator and therefore  $\lambda_n=3H n$ and 

\begin{eqnarray}
\Psi^{(0)}_n(v_\zeta)&=&\sqrt{\frac{3H}{\pi}}\frac{1}{2^n n!}H_n\left(\sqrt{3H/D_\zeta} v_\zeta\right) {\rm exp}\left(-3 H v_\zeta^2/D_\zeta\right)\, ,\nonumber\\
\Psi^{(0)\dagger}_n(v_\zeta)&=&H_n\left(\sqrt{3H/D_\zeta} v_\zeta\right)\, ,
\end{eqnarray}
where $H_n(x)$ are the Hermite polynomials. 
We now proceed computing the operators $L_{0,0}$, $L_{0,n}$ and $L_{0,n}$. The operator $L_{0,0}$ 
is clearly vanishing. To compute $L_{0,n}$, we make use of the fact that $\sqrt{3H/D_\zeta} v_\zeta=1/2 \Psi_1^{(0)\dagger}=1/2 H_1\left(\sqrt{3H/D_\zeta} v_\zeta\right)$. Using the orthonormality condition
(\ref{ort}), this gives

\be
L_{0,n}=-\frac{\partial}{\partial\zeta}\sqrt{\frac{D_\zeta}{12 H}}\,\delta_{1,n}\, .
\ee
This means that we need only to compute $L_{1,0}$, which can be done by exploiting the fact that
$6H/D_\zeta v_\zeta^2=1/2\Psi_2^{(0)\dagger}+1$, giving

\be
L_{1,0}=-\frac{\partial}{\partial\zeta}\sqrt{\frac{D_\zeta}{3 H}}\, .
\ee
Therefore the equation satisfied by the slow variable $\zeta$, from (\ref{ppp}) is,  the KM
equation

\be
\frac{\partial}{\partial t}{P}(\zeta,t)=\frac{D_\zeta}{18 H^2}\frac{\partial^2}{\partial\zeta^2}P(\zeta,t)+\cdots\, ,
\ee
which gives $
\langle\zeta^2\rangle =(D_\zeta/9 H^2)t$ 
and  coincides with (\ref{o}) as expected. Notice that the same result  can be 
also obtained in an alternative way. From Eq. (\ref{q}) with quadratic potential $V(v_\zeta)$, we may
derive the set of equations

\begin{eqnarray}
\frac{\partial}{\partial t}\langle \zeta^2\rangle&=&2\langle\zeta v_\zeta\rangle\, ,\nonumber\\
\frac{\partial}{\partial t}\langle \zeta v_\zeta\rangle&=&\langle v^2_\zeta\rangle
-3H\langle\zeta v_\zeta\rangle\, ,\nonumber\\
\frac{\partial}{\partial t}\langle v^2_\zeta\rangle&=&-6H\langle v^2_\zeta\rangle
+D_\zeta
\, .
\end{eqnarray}
At times larger than a few Hubble times, the correlators involving $v_\zeta$ decay promptly
to their equilibrium values

\be
\langle\zeta v_\zeta\rangle=\frac{1}{3H}\langle v^2_\zeta\rangle=\frac{D_\zeta}{18H^2}\, ,
\ee
resulting in 
\be
\frac{\partial}{\partial t}\langle \zeta^2\rangle=\frac{D_\zeta}{9H^2}
\, ,
\ee
whose solution reproduces once more (\ref{o}).

\section{Coupling to an en extra scalar field}
\noindent
As further  example, we consider the inflationary theory originally studied in Ref. \cite{loops1} where
an inflaton with a standard kinetic term is rolling down a flat potential and is interacting
gravitationally with $N$ scalar fields $\sigma_n$. We will consider from now on
only one single scalar field as our conclusions  do not change if the number of fields is larger than one.
We will also suppose that the scalar field has a potential $V(\sigma)$ and we will focus on its
consequences on the probability for the comoving curvature perturbation as done in Ref. \cite{woo} and, partially, in Ref.  \cite{loops6}.
The action we are concerned with to extract the IR behaviour therefore becomes

\be
\label{actionsigma}
S=\int d^4x a^3\left[\epsilon\mpl^2\dot\zeta^2+
\frac{1}{2}\dot{\sigma}^2\left(1+3\zeta-\frac{\dot\zeta}{H}\right)
-\left(1+\frac{1}{H a^3}\frac{d}{{\rm d}t}
\left(a^3\zeta\right)\right)V(\sigma)
+\cdots\right]\ ,
\ee
where the dots stand for further interaction between $\zeta$ and $\sigma$ that we disregard and the
interaction between $\zeta$ and the scalar potential $V(\sigma)$ acquires that particular form once 
we take into account that the linear $\zeta$ dependence in the metric,  $ds^2=(1+\dot\zeta/H)^2 {\rm d}t^2-a^3(1+3\zeta)d{\bf x}^2$.
Following again the rules dictated by \cite{TW}, the IR part of the comoving curvature perturbation
and the field $\sigma$ satisfy the equations

\begin{eqnarray}
\label{ii}
3H\dot\zeta-\frac{1}{2H\epsilon\mpl^2}\dot\sigma\left(3H\dot\sigma+V'(\sigma)\right)&=&3H\eta_\zeta\, ,\nonumber\\
\ddot{\sigma}+3H\dot\sigma+V'(\sigma)&=&\eta_\sigma\, ,
\end{eqnarray}
where we have introduced the Gaussian and Markovian noise $\eta_\sigma$ with correlator

\be
\langle \eta_\sigma({\bf x},t)\eta_\sigma({\bf x},t')\rangle=\frac{H^3}{4\pi^2}\delta_D(t-t')\ .
\ee
In Eq. (\ref{ii})  we have neglected for simplicity the term with two derivatives with respect to time
of the comoving curvature perturbation; furthermore,   the $\zeta$ dependence in the equation of $\sigma$ disappear at the order we are working upon the substitutions dictated by Ref. \cite{TW}. Introducing  the variable
$v_\sigma=\dot\sigma$, Eq. (\ref{ii}) can be rewritten as 

\begin{eqnarray}
\label{iii}
\dot\zeta-\frac{1}{6H^2\epsilon\mpl^2}v_\sigma\left(3Hv_\sigma+V'(\sigma)\right)&=&\eta_\zeta\, ,\nonumber\\
\dot\sigma&=&v_\sigma\, ,\nonumber\\
\dot{v}_\sigma+3H v_\sigma+V'(\sigma)&=&3H\eta_\sigma\, .
\end{eqnarray}
The equation for the probability $P(\zeta,\sigma,v_\sigma,t)$ then reads

\be
\label{iu}
\frac{\partial}{\partial t} P=-\frac{(3Hv_\sigma+V'(\sigma))}{6H^2\epsilon\mpl^2}v_\sigma\frac{\partial}{\partial \zeta} P
+\frac{D_\zeta}{18 H^2}\frac{\partial^2}{\partial \zeta^2} P-v_\sigma\frac{\partial}{\partial \sigma} P+
\frac{\partial}{\partial v_\sigma}\left(3H v_\sigma+V'(\sigma)+\frac{9H^5}{8\pi^2}\frac{\partial}{\partial v_\sigma}\right)P\, ,
\ee
with initial condition

\be
P(\zeta,\sigma,v_\sigma,0)=\delta_D(\zeta)\delta_D(\sigma)\delta_D(v_\sigma)\, .
\ee
The operator

\be
L_{\sigma v_\sigma}=-v_\sigma\frac{\partial}{\partial \sigma} P+
\frac{\partial}{\partial v_\sigma}\left(3H v_\sigma+V'(\sigma)+\frac{9H^5}{8\pi^2}\frac{\partial}{\partial v_\sigma}\right)\, 
\ee
can be written as

\be
L_{\sigma v_\sigma}=-{\rm exp}\left(-\frac{2}{3}\pi^2\frac{v_\sigma^2}{H^4}-\frac{4}{3}\pi^2
\frac{V(\sigma)}{H^4}\right)\left(3H b^\dagger b+bD+b^\dagger \hat D\right)
{\rm exp}\left(\frac{2}{3}\pi^2\frac{v_\sigma^2}{H^4}+\frac{4}{3}\pi^2
\frac{V(\sigma)}{H^4}\right)\, ,
\ee
where

\begin{eqnarray}
b&=&\frac{3H^2}{2\sqrt{6}\pi}\frac{\partial}{\partial v_\sigma}+\frac{\pi\sqrt{6}}{3H^2}v_\sigma\,,\nonumber\\
b^\dagger&=&-\frac{3H^2}{2\sqrt{6}\pi}\frac{\partial}{\partial v_\sigma}+\frac{\pi\sqrt{6}}{3 H^2}v_\sigma\,,\nonumber\\
D&=&\frac{3H^2}{2\sqrt{6}\pi}\frac{\partial}{\partial \sigma}-\frac{\pi\sqrt{6}}{3H^2}V'(\sigma)\,,\nonumber\\
\hat D&=&-D^\dagger=\frac{3H^2}{2\sqrt{6}\pi}\frac{\partial}{\partial \sigma}+\frac{\pi\sqrt{6}}{3H^2}V'(\sigma)\,.
\end{eqnarray}
The normalized eigenfunctions $\Psi_n(v_\sigma)$ of the operator $b^\dagger b$, {\it i.e.} $b^\dagger b
\Psi_n(v_\sigma)=n\Psi_n(v_\sigma)$ with $n\geq 0$, are the eigenfunctions of the standard harmonic oscillator
and can expressed in terms of the Hermite polynomials. They can be used to expand the probability
as

\be
\label{xc}
P(\zeta,\sigma,v_\sigma,t)=\Psi_0(v_\sigma){\rm exp}\left(-4\pi^2 V(\sigma)/3H^4\right)
\sum_{n=0}^\infty P_n(\zeta,\sigma,t)\Psi_n(v_\sigma)\, ,
\ee
where 

\be
\Psi_0(v_\sigma)=\left(-2\pi^2 v_\sigma^2/3H^4\right)/\sqrt{3H^2/8\pi\sqrt{3\pi}}\, .
\ee
The functions $P_n(\zeta,\sigma,t)$ satisfy the equations

\begin{eqnarray}
\frac{\partial}{\partial t}P_n&=&-\frac{V'(\sigma)}{6H^2\epsilon\mpl^2}\frac{3H^2}{2\pi\sqrt{6}}
\frac{\partial}{\partial \zeta}\left(\sqrt{n}P_{n-1}+\sqrt{n+1}P_{n+1}\right)\nonumber\\
&-&\frac{3H}{6H^2\epsilon\mpl^2}\frac{9H^4}{24\pi^2}
\frac{\partial}{\partial \zeta}\left(\sqrt{n(n-1)}P_{n-2}+n P_n+\sqrt{(n+1)(n+2)}P_{n+2}+(n+1)P_n\right)\nonumber\\
&+&\frac{D_\zeta}{18 H^2}\frac{\partial^2}{\partial \zeta^2}P_n
\nonumber\\
&-&\sqrt{n}\hat D P_{n-1}-3HnP_n-\sqrt{n+1}D P_{n+1}\, .
\end{eqnarray}
They split into
\begin{eqnarray}
\frac{\partial}{\partial t}P_0&=&-\frac{V'(\sigma)}{6H^2\epsilon\mpl^2}\frac{3H^2}{2\pi\sqrt{6}}
\frac{\partial}{\partial \zeta}P_1
-\frac{3H}{6H^2\epsilon\mpl^2}\frac{9H^4}{24\pi^2}
\frac{\partial}{\partial \zeta}\left(P_0+\sqrt{2}P_2\right)\nonumber\\
&+&\frac{D_\zeta}{18 H^2}\frac{\partial^2}{\partial \zeta^2}P_0
-D P_{1}+\cdots\, .
\end{eqnarray}
and
\begin{eqnarray}
3HP_1&=& -\frac{V'(\sigma)}{6H^2\epsilon\mpl^2}\frac{3H^2}{2\pi\sqrt{6}}
\frac{\partial}{\partial \zeta}\left(P_{0}+\sqrt{2}P_2\right)-\hat{D} P_{0}+\cdots\, ,\nonumber\\
6HP_2&=&-\frac{3H}{6H^2\epsilon\mpl^2}\frac{9H^4}{24\pi^2}
\frac{\partial}{\partial \zeta}\sqrt{2}P_0+\cdots\, ,
\end{eqnarray}
where we have supposed that $t\gg H^{-1}$ so that the probabilities $P_n$ with $n\geq 1$ have asymptotically reached stationary solutions. It is easy to convince oneself that 
the   $\sigma$-dependence of $P_0$ is of the stationary type
\be
P_0(\zeta,\sigma,t)=P_0(\zeta,t){\cal N}_0{\rm exp}\left(-4\pi^2 V(\sigma)/3H^4\right)\, ,
\ee
where ${\cal N}_0$ is a normalization factor. The expression above satisfies $\hat{D}P_0=0$. 
Due to the orthonormality conditions for the $\Psi(v_\sigma)$, the probability $P(\zeta,t)$ coincides
with $P_0(\zeta,t)$ and we 
get the  following KM equation for $P(\zeta,t)$

\begin{eqnarray}
\frac{\partial}{\partial t}P(\zeta,t)&=&\frac{1}{2}D_\zeta^{\rm eff}\frac{\partial^2}{\partial \zeta^2}P(\zeta,t)+\frac{1}{3!}\,S^{\rm eff}_\zeta\frac{\partial^3}{\partial \zeta^3}P(\zeta,t)+\cdots\, ,
\nonumber\\
\frac{1}{2}D_\zeta^{\rm eff}&=&\frac{D_\zeta}{18 H^2}+\frac{1}{3H}\left(\frac{1}{4\pi\sqrt{6}\epsilon\mpl^2}\right)^2\langle\left(V'(\sigma)\right)^2\rangle+\frac{\sqrt{2}}{6H}\left(\frac{9H^3}{48\pi^2\epsilon\mpl^2}\right)^2\, ,\nonumber\\
\frac{1}{3!}\,S^{\rm eff}&=&\left(\frac{1}{12\epsilon\pi^2\mpl^2}\right)^2\left(\frac{H}{24\epsilon\mpl^2}\right)
\langle\left(V'(\sigma)\right)^2\rangle\, .
\end{eqnarray}
where one recognizes that the corrections to the tree-level diffusion coefficient originate from the 
one-loop contribution to  the two-point correlator of $\zeta$ obtained by exchanging the $\sigma$  through the potential $V(\sigma)$ (second term in the right hand side of of the middle line)
and $\dot\sigma=v_\sigma$ (last term in the right hand side of of the middle line) degrees of freedom. The term proportional to one derivative with respect to $\zeta$ is absent once it is insured that the expectation value of $\zeta$ vanishes at all orders. The fact that $\langle\zeta^2\rangle \sim t$ (again neglecting the time change of the slow-roll variables) is due  to the fast dynamics of the $\sigma$ field and  $v_\sigma$ which can be integrated out as they reach equilibrium values on a time scale of the Hubble time.

\section{The case of interactions dominated by slow-roll terms}
\noindent
As last example,  we will be concerned with the case in which inflation is driven by a single scalar field  with 
canonical kinetic term. In such a case the  self-interactions of the comoving curvature perturbation are dominated by  terms which are suppressed by 
slow-roll parameters. To (considerably) simplify the matter we work in the limit in which
the slow-roll parameter $|\eta|=|\dot{\epsilon}/\epsilon H|\gg |\epsilon|$.  This happens if the model
of inflation is, for instance, an hybrid model with potential $V(\phi)=V_0+m^2\phi^2/2$; in this
case $\eta$ is constant and equal to $(m^2/3 H_0^2)=(m^2\mpl^2/V_0)$, while $\epsilon=\phi_{\rm f}{\rm exp}(-\eta N)$ 
(where $N$ is the number of  e-folds till the end of inflation and $\phi_{\rm f}$ is the value of the 
inflaton field at the waterfall transition ending inflation) is exponentially suppresed \cite{lr}.
The curvature perturbation action simplifies to \cite{maldacena}

\be
\label{eq:action1}
S=\mpl^2\int d^4x\left[a^3\epsilon\left(\dot\zeta^2-\frac{1}{a^2}(\nabla\zeta)^2\right)
+\frac{1}{2}a^3\epsilon\,\dot{\eta}\,\zeta^2\,\dot{\zeta}
+\frac{1}{2}\eta\,\zeta^2\frac{d}{{\rm d}t}\left(a^3\epsilon\,\dot{\zeta}\right)-\frac{1}{2}\frac{d}{{\rm d}t}\left(a^3\epsilon\,\eta\,
\dot\zeta\zeta^2\right)+\cdots\right]\, .
\ee
Applying the rules of Ref. \cite{TW} and working to order ${\cal O}(\eta)$ we find the equation for the
IR part of $\zeta$

\be
\ddot\zeta+3H\dot\zeta-3H\eta\,\zeta\,\dot\zeta=3H\eta_\zeta\, .
\ee
With the usual position $\dot\zeta=v_\zeta$, the previous equation can be written as

\begin{eqnarray}
\label{lo}
\dot\zeta&=&v_\zeta\, ,\nonumber\\
\dot{v}_\zeta+3Hv_\zeta(1-\,\eta\,\zeta)&=&3H\eta_\zeta\, .
\end{eqnarray}
At this stage we might proceed as in the previous section by integrating out the fast variable
$v_\zeta$ after having found the eigenfunctions and eigenvalues of the corresponding operator
$L_{v_\zeta}$.  We take this time a short-cut and eliminate $v_\zeta$ as follows.
In  Eq. (\ref{lo}) we neglect the time derivative of $v_\zeta$ and solve for the latter

\be
v_\zeta\simeq \frac{\eta_\zeta}{1-\eta\,\zeta}\, ,
\ee
which gives

\be
\label{ax}
\frac{\dot\zeta}{1+\eta\zeta}\simeq \eta_\zeta\, .
\ee
Instead of finding the corresponding KM equation, we follow this time an alternative
road. We integrate directly Eq. (\ref{ax}) after having changed the variable of integration from $t$ to $N=\int^t H(t') {\rm d}t'$

\be
\zeta(N)=\frac{1}{\eta}\left[{\rm exp}\left(\eta\int_0^N {\rm d}N'\,\frac{\eta_\zeta}{H}\right)-1\right]\, .
\ee
Suppose we are interested in the variance computed at some e-fold $N$ which corresponds to a scale
inside our current comoving Hubble radius. Therefore we can write $N=(N-N_{60})+N_{60}$ where
now $N_{60}$ corresponds to the number of e-folds computed from the beginning of inflation till the moment when the wavelengh corresponding to our current Hubble radius exited the horizon during inflation. We can therefore write

\bea
\zeta(N)&=&\frac{1}{\eta}\left[{\rm exp}\left(\eta\int_0^{N_{60}} {\rm d}N'\,\frac{\eta_\zeta}{H}+
\eta\int^{N}_{N_{60}} {\rm d}N'\,\frac{\eta_\zeta}{H}
\right)-1\right]\nonumber\\
&\simeq&
\frac{1}{\eta}\left[{\rm exp}\left(\eta\int_0^{N_{60}} {\rm d}N'\,\frac{\eta_\zeta}{H}\right)\left(
1+\eta\int^{N}_{N_{60}} {\rm d}N'\,\frac{\eta_\zeta}{H}
\right)-1\right]
\, .
\eea
Due to the fact that the noise is white and therefore there is no cross-correlation between noises evaluated at different times, the variance becomes

\be
\langle\zeta^2\rangle\simeq \Big\langle{\rm exp}\left(\eta\int_0^{N_{60}} {\rm d}N'\,\frac{\eta_\zeta}{H}\right)
{\rm exp}\left(\eta\int_0^{N_{60}} {\rm d}N''\,\frac{\eta_\zeta}{H}\right)
\Big\rangle\,\Big\langle \int^{N}_{N_{60}} {\rm d}N'\,\frac{\eta_\zeta}{H}\int^{N}_{N_{60}} {\rm d}N''\,\frac{\eta_\zeta}{H}\Big\rangle\, .
\ee
As the noise $\eta_\zeta$ is Gaussian and for any Gaussian quantity ${\cal O}$ the property
$\langle {\rm exp}\, {\cal O}\rangle={\rm exp}\,2\langle {\cal O}^2\rangle$ holds, we obtain

\be
\label{xc}
\langle\zeta^2(N)\rangle\simeq {\rm exp}\left(\frac{\left(n_\zeta-1\right)^2}{2}
\langle\zeta^2(N_{60})\rangle\right)\,\langle\zeta^2(N-N_{60})\rangle\,,
\ee
where we have traded the slow-roll parameter  $\eta$ with the spectral index of the comoving curvature perturbation $(n_\zeta-1)=2\eta$. This result coincides with the semi-classical result obtained in Ref. \cite{loops7} once one neglects there the contribution from the running of the spectral index which
in our example is $\ll\eta^2$. Indeed, in Ref. \cite{loops7} the power spectrum for modes inside our current Hubble radius $H_0^{-1}$ receives IR contributions  once it is averaged by changing the   location of the sphere 
with radius $\sim H_0^{-1}$
inside the superlarge box (corresponding to the total number of e-folds $N$). This is not surprising since the  power spectrum computed in our Hubble volume, but averaged inside the superlarge box,  is nothing else that the power spectrum inside the superlarge box. The latter is affected  by large  IR divergences, but is not directly observable.

Notice that the result can be obtained by writing the corresponding 
approximated equation governing the probability $P(\zeta,t)$  \cite{haken1}. The adiabatic elimination of the fast variable $v_\zeta$  adopted in the previous sections
indeed gives

\be
\label{nm}
\frac{\partial }{\partial N}P(\zeta,N)=\frac{D_\zeta}{18 H^3}\frac{\partial}{\partial\zeta}\left[
(1+\eta \zeta)\frac{\partial}{\partial\zeta} (1+\eta \zeta)
 P(\zeta,N)\right]=
\frac{D_\zeta}{18 H^3} \frac{\partial^2}{\partial X^2}P(X,N)
 \, , 
\ee
where ${\rm d} X={\rm d} \zeta/(1+\eta \zeta)$. Recalling that $P(X,N){\rm d}X=P(\zeta,N){\rm d}\zeta$
and 
to the order in the slow-roll parameters we are working at, we obtain
that 

\be
\langle\zeta^2(N)\rangle\simeq \int\,{\rm d}X \frac{1}{\eta^2}\left[{\rm exp}\left(\eta X\right)-1\right]^2\,P(X,N)\, .
\ee
As $P(X,N)$ is approximately the Gaussian function (after a change variable for $N$ to incorporate
the time dependence of the diffusion coefficient), one reproduces the result (\ref{xc}).

\section{Conclusions}
\noindent
The study of interacting fields in cosmological backgrounds is a field which has attracted 
recently a lot of attention as future experiments will measure
 correlation functions of cosmological fluctuation
variables with large accuracy. Cosmological correlators will be affected by ultraviolet divergences, 
as in quantum field theory, and by infrared divergences from modes which left the horizon at some time earlier
than the relevent observable mode. The stochastic approach offers a technique to study
such infrared effects. 
The infrared part of the curvature perturbation  performs a random walk under the action of the stochastic noise
generated by those modes which continuously leave the Hubble radius. 
The equation governing the probability distribution of the comoving curvature
perturbation   to acquire a given value  at a given spatial point  and time
is the  Kramers-Moyal equation,  which generalizes the more standard (at least in cosmology) 
diffusive Fokker-Planck equation. 
In section 2 and 3 we have for simplicity ignored slow-roll suppressed interactions, noting that the interactions leading to large IR effects are slow-roll suppressed and hence ignored in this limit. In section 4, we have shown that in the specific example, considering the interactions with a test scalar field, 
the late time behaviour of the effects on the two-point correlator 
of the curvature perturbation $\zeta$ is consistent with the fact that $\zeta$ is conserved
on superhorizon scales. The infrared effects considered in the example do not alter this property, as expected. In section 5, we considered a specific example of slow-roll inflation, and were able to reproduce the results of the semi-classical relations in \cite{loops7}.
Our work is far from being completed. First of all, we have derived in its full generality
the Kramers-Moyal equation from a path-integral approach and shown how to alternatively derive
such an equation through the adiabatic elimination of fast variables in a selected set of examples. 
One would certainly like to generalize and systematize this procedure. In particular, the role
of self-interaction terms where spatial gradients of the curvature perturbation are present
have not been addressed in this paper. This is because stochastic approach turns out to be surprisingly technically cumbersome when including derivative interactions. But although a simpler alternative semi-classical method has been developed in \cite{loops7} (see also \cite{ir6,ir7} for further discussions) for computing IR corrections to correlation functions of cosmological perturbations, the stochastic approach may still be useful for understanding some alternative or complementary questions.  
Furthermore, we have 
taken for granted that the stochastic approach leads to a resummation of the leading
infrared effects. While results in the literature based on the study of self-interacting scalar fields
in a de Sitter stage support  this expectation, a more formal derivation for the case of the curvature
perturbation and its interactions is desirable.

\section*{Acknowledgments}
\noindent
We would like to thank Robert Brandenberger discussions and comments on the draft, and Steven B. Giddings for discussions on IR issues in de Sitter.

\bibliographystyle{JHEP}

\end{document}